\documentclass[12pt]{article}
\pdfoutput=1
\usepackage{amsmath,amssymb,amsfonts,color,graphicx,color,cite}
\usepackage{mciteplus}
\usepackage{slashed}
\usepackage[bottom]{footmisc}
\usepackage{float}
\usepackage[plainpages=false,colorlinks,linkcolor=blue,citecolor=blue,urlcolor=blue]{hyperref}

\usepackage{booktabs}

\hypersetup{pdfauthor={Bastian Feigl, Heidi Rzehak and Dieter Zeppenfeld},
            pdftitle={New Physics backgrounds to the H->WW search at the LHC?}}

\mciteSetMidEndSepPunct{;\newline}{.}{\relax}

\evensidemargin \oddsidemargin
\usepackage{sectsty}
\sectionfont{\Large}
\subsectionfont{\large}
\subsubsectionfont{\large}

\allowdisplaybreaks

\begin{document}
\thispagestyle{empty}

\def\thefootnote{\fnsymbol{footnote}}
\def\pslash#1{{\setbox0=\hbox{$#1$}
  \rlap{\ifdim\wd0>.18em\kern.18\wd0\else\kern.18\wd0\fi /}#1}}

\begin{flushright}
KA--TP--21--2012\\
CERN-PH-TH/2012-131
\end{flushright}

\vspace{0.5cm}

\begin{center}

{\large\sc {\bf New Physics backgrounds to the $H\to WW$ search at the LHC?}}

\vspace{1cm}

{\sc
B.~Feigl$^{1}$%
\footnote{email: bastian.feigl @ kit.edu}%
, H.~Rzehak$^{2}$\footnote{On leave from:
  Albert-Ludwigs-Universit\"at Freiburg, Physikalisches Institut,
  Freiburg, Germany.}%
\footnote{email: heidi.rzehak @ cern.ch}%
~and D.~Zeppenfeld$^{1}$%
\footnote{email: dieter.zeppenfeld @ kit.edu}
}

\vspace*{.7cm}

{$^1$Institut f\"ur Theoretische Physik, Karlsruher Institut f\"ur Technologie \\
D--76128 Karlsruhe, Germany

\vspace*{0.1cm}

$^2$ CERN, PH-TH, 1211 Geneva 23, Switzerland
}

\end{center}

\vspace*{0.1cm}

\begin{abstract}
\noindent The searches for $H\to WW$ events at the LHC use data driven
techniques for estimating the $q\bar q\to WW$ background, by normalizing the
background cross section to data in a control region. We investigate the
possibility that new physics sources which mainly contribute to the control
region lead to an overestimate of Standard Model backgrounds to the Higgs boson signal
and, thus, to an underestimate of the $H\to WW$ signal. A supersymmetric
scenario with heavy squarks and gluinos but charginos in 
the 200 to 300 GeV region and somewhat lighter sleptons can lead to such a situation.
\end{abstract}


\def\thefootnote{\arabic{footnote}}
\setcounter{page}{0}
\setcounter{footnote}{0}

\newpage

\section{Introduction}
Among the Higgs search channels at the LHC, the decay into weak
boson pairs is particularly important, because the tree level couplings of the
Higgs boson to $WW$ and $ZZ$ allow to positively identify a scalar state as
being associated with the vacuum expectation value which breaks the
electroweak gauge symmetry. 
A Higgs boson produced via gluon fusion with subsequent decay into leptonically 
decaying $W$ bosons is especially searched for, at the present time, because
this channel is expected to produce the largest sample of $H\to VV$ events at the LHC. 
However, for a Higgs boson mass around 125~GeV, it does suffer from
substantial backgrounds, the dominant one being  $q\bar q\to W^+W^-$, followed by 
leptonic decay of the W~bosons. In the present experimental analyses, this
contribution to the background is estimated with data driven 
methods~\cite{lhchiggsgroup, atlasHinWW, cmsHinWW}. 
A control region of the phase space, namely events at large dilepton invariant
mass $m_{\ell\ell}$, is defined where no signal 
events are to be expected. The size of the background is measured
in this control region and the corresponding value in the signal region, at
small $m_{\ell\ell}$, is 
then extrapolated using the shape of the distribution determined by a Monte
Carlo simulation for Standard Model (SM) W~pair production. 

In this paper we investigate whether new physics contributions can seriously
compromise this data driven background determination. Specifically, can new
physics events, which typically are hard and thus tend to preferentially
populate the large $m_{\ell\ell}$ region,  significantly enhance the size of the
background measured in the control region, while contributing relatively less
than the SM W~pair production to the signal region? This would lead to an
over-estimate of the background in the signal region and thus to a 
significant underestimate of the Higgs signal. Since the presently measured
$H\to WW$ rates are indeed below SM expectations, for 
$m_H\approx 125$~GeV~\cite{atlasHinWW, cmsHinWW},  such  scenario deserves
serious investigation.

We will focus on processes arising within the Minimal Supersymmetric 
Standard Model (MSSM) as an example for processes induced by physics beyond
the Standard Model (BSM physics).
SUSY processes contributing as background have been discussed before with respect to calibration 
processes \cite{Baer} and to Higgs boson searches within vector boson fusion 
where the Higgs boson decays into two tau~leptons or two W~bosons \cite{SUSYbackVBF}. Also, 
in \cite{electroweakinos} SUSY processes as a possible background in the signal region of Higgs boson searches 
have been considered.

\section{Scenario and analysis setup}
\label{scenario}

Within the MSSM, processes that can contribute significantly to the signature of W boson pair production
involve the production of charginos, sleptons and neutralinos and therefore the parameters that influence
their masses are the most relevant ones.

Our scenario is based on the light slepton scenario described in \cite{SUSYbackVBF}, which is tuned
for sizable chargino and slepton pair production. The tau slepton masses are chosen to be above the light
chargino mass while the sleptons of the first two generations are lighter than the light chargino.
As a consequence, the main decay channels of the charginos are via a slepton of the first two 
generations and a corresponding lepton, where the slepton decays directly into a lepton and the lightest neutralino (which is 
assumed to be the lightest supersymmetric particle (LSP)).
We increase the chargino mass by modifying the soft SUSY breaking parameters $M_2$ and $m_{H_u}$ to 
be in agreement with the ATLAS $2\;\textrm{fb}^{-1}$ trilepton search \cite{atlascharginos}.
However our scenario shows strong tensions with the CMS trilepton 
analysis%
\footnote{A possibility to weaken
this tension is to allow a decay of the light chargino and the next-to-lightest neutralino into all three lepton flavors in 
equal parts. However this reduces the effect on the correction factor~C by roughly 35-40\%.} 
\cite{cmscharginos1,*cmscharginos2} 
which uses the full $5\;\textrm{fb}^{-1}$ dataset of 2011, if the predicted neutralino properties
are taken literally. Since production processes with next-to-lightest neutralinos are not important for our analysis (see below), 
we ignore this tension having in 
mind alternative models that avoid the visible leptonic signature of the 
$\chi_2^0$.%
\,\footnote{This can be achieved for example with a scenario (typically beyond the MSSM) where the 
next-to-lightest neutralino is much heavier than the light chargino.}
The masses of squarks of the first two generations and of the gluino are not important for the following study
and are set to high values. 
The parameters in the Higgs boson sector of the MSSM are assumed to have values which 
lead to a Standard Model type Higgs boson with mass of $124.7$~GeV, which is 
within the experimentally allowed Higgs boson mass range and in the mass area where some 
experimental hints of a Higgs boson have 
been observed~\cite{atlashiggscomb1,*atlashiggscomb2,cmshiggscomb1,*cmshiggscomb2}.
Starting from this ``base'' scenario we vary the parameters that control the 
chargino, 
slepton and lightest neutralino masses to study their influence on the production and decay of the 
SUSY particles and their contribution to the W pair production signature.
A scenario with 25\% higher LSP and 40\% higher slepton masses is discussed in more detail.
The full parameter settings of the SUSY scenarios can be found in Appendix~\ref{parameter}.
The following SUSY production processes give the dominant contributions to the signature of
the W boson pair production:
\begin{eqnarray}
q \, \bar{q} \rightarrow \chi_1^+ \; \chi_1^-
&\rightarrow& \ell^+\;\ell^- + \slashed{p}_T \label{ppx1x1}\\
q \, \bar{q} \rightarrow \widetilde{\ell}^+ \; \widetilde{\ell}^-
&\rightarrow& \ell^+\;\ell^- + \slashed{p}_T \label{ppslsl}\\
q \, \bar{q} \rightarrow \chi_1^\pm \; \chi_2^0
&\rightarrow& \ell_1^+\;\ell_2^+\;\ell_2^- + \slashed{p}_T ~. \label{ppx1n2}
\end{eqnarray}
Chargino pair production gives the by far largest contribution, followed by the
slepton pair production processes. The production of a next-to-lightest neutralino and a chargino
gives only a very small contribution to the W boson pair production
signature, as it produces mostly three leptons and the third lepton can be tagged quite well. 
Therefore this channel is not crucial for our results and general BSM scenarios without such a 
trilepton source can evade detection and still show the same
behaviour concerning the $WW$ background estimation.

It should be emphasized that the important feature of the scenarios is the existence 
of new physics particles which decay into electrons, muons and invisible particles, 
which is the same particle content in the final state as the one of the W pair production 
with subsequent leptonic decay. In that respect we consider our MSSM scenarios as an 
example for BSM physics.

The event generation for our analysis is done using {\tt Herwig++ 2.5.2}~\cite{Herwig,*Herwig25}
at parton level, including parton shower. Within {\tt Herwig++}, the W pair production is generated
at next-to-leading order QCD within the POWHEG framework~\cite{powhegWW}. The SUSY particle pairs 
are generated at leading order \cite{herwigBSM1,*herwigBSM2}. Their cross section is multiplied
with an appropriate K-factor ($K=1.2$), which we obtain from 
{\tt Prospino2}~\cite{prospinocharginos}.
Spin correlations within production and decay in {\tt Herwig++} are included as described in \cite{Richardson:2001df}.
A comparison with a combination of {\tt MadGraph 5.1.3}~\cite{MadGraph5,*Madgraph4,*MadgraphNew}
and {\tt Pythia~6.4}~\cite{pythia} led to comparable results.

Both ATLAS~\cite{atlasHinWW} and CMS~\cite{cmsHinWW} have presented a study of the 
$H \rightarrow WW \rightarrow 2\ell 2\nu$ channel with the full data set of 2011.
We largely use the cuts and methods from the ATLAS analysis, because they show
distributions in the transverse mass of the W boson pair~\cite{transversemass}
\begin{equation}
  m_T = \sqrt{(E_T^{\ell\ell}+E_T^{miss})^2-|\mathbf{p}_T^{\ell\ell}+\mathbf{p}_T^{miss}|^2},
        \quad \textrm{with} \; E_T^{\ell\ell} = \sqrt{|\mathbf{p}_T^{\ell\ell}|^2+ m_{\ell\ell}^2},\; 
        E_T^{miss} = |\mathbf{p}_T^{miss}|
\end{equation}
in the signal and control regions up to quite high values in $m_T$. 
This gives an opportunity to check if the effects of BSM physics could be
identified in the experiment as an excess of particularly hard events.
CMS performs a similar analysis, but they only show distributions of the
invariant lepton pair mass $m_{\ell\ell}$ and the azimuthal angle $\Delta \phi_{\ell\ell}$ between the leptons. 
Those distributions have turned out to be less illuminating for our study. However, as the CMS cut selection 
is similar to the ATLAS one, our results should hold qualitatively for CMS as well. 

Our analysis is carried out for the LHC operating at a center-of-mass energy
of 7 TeV.
We use the {\tt CT10} parton distribution functions~\cite{CT10} for the
POWHEG event samples and {\tt cteq6l1} pdfs for the leading order SUSY calculations.
As renormalization and factorization scales we use the invariant mass of the W boson pair
or the SUSY particle pair.
For the basic event selection we require two oppositely charged leptons $\ell$ (electrons or muons),
where the harder lepton with respect to transverse momentum $p_T$ is labeled $\ell_1$, the softer one $\ell_2$.
The following cuts are applied, largely taken from~\cite{atlasHinWW}:
\begin{equation}
\begin{array}{rclcrcl}
  p_{T,\ell_1} &>& 25\,\textrm{GeV} & \quad &
  p_{T,\ell_2} &>& 15\,\textrm{GeV}  \\
  m_{ee\,(\mu\mu)} &>& 12\,\textrm{GeV} & \quad &
  m_{e\mu} &>& 10\,\textrm{GeV} \\
  |m_{ee\,(\mu\mu)}-M_Z| &>& 15\,\textrm{GeV} & \quad &
  |\eta_\ell| &<& 2.5 ~.
\end{array}
\label{basecuts}
\end{equation}
The events are categorized according to the number of visible jets.
Jets are clustered using the anti-$k_t$ algorithm~\cite{antikt} with distance parameter $R=0.4$ and
the following requirements on rapidity $\eta_j$ and $p_{T,j}$:
\begin{equation}
|\eta_j| < 4.5 \quad\quad p_{T,j}>25\,\textrm{GeV}~.
\end{equation}
Leptons that are within the R-separation $\Delta R < 0.3$ of a jet are counted as part of the jet.
For QCD background suppression in the $H\rightarrow WW$ analysis the LHC experiments use
the quantity $E_{T,rel}^{miss}=E_{T}^{miss} \cdot \textrm{sin} \, \textrm{min}(\Delta\phi,\,\frac{\pi}{2})$, 
where $E_{T}^{miss}$ is the missing transverse energy of the event and $\Delta\phi$ is the azimuthal angle
between the $E_{T}^{miss}$ vector and the nearest lepton or jet with $p_T > 25\,\textrm{GeV}$.
The requirement is
\begin{equation}
  E_{T,rel}^{miss} > 45\,\textrm{GeV for $\ell\ell=ee\,(\mu\mu)$} \quad\quad  E_{T,rel}^{miss} > 25\,\textrm{GeV for 
  $\ell\ell=e\mu$}~.
\label{etmissrelcut}
\end{equation}
The spin-0 nature of the Higgs boson is exploited by demanding~\cite{philldreiner}
\begin{equation}
  \Delta \phi_{\ell\ell} < 1.8 ~.
\label{phillcut}
\end{equation}
The signal region in the 0-jet channel is furthermore restricted in the dilepton transverse momentum 
$p_T^{\ell\ell}$
\begin{equation}
  p_T^{ee\,(\mu\mu)} > 45\,\textrm{GeV} \quad\quad p_T^{e\mu} > 30\,\textrm{GeV}
\label{ptllcut}
\end{equation}
and~\cite{barger}
\begin{equation}
  m_{\ell\ell} < 50 \,\textrm{GeV} ~.
\label{mllmaxcut}
\end{equation}
For the 1-jet channel ATLAS uses cuts on the vectorial sum of the $\mathbf{p}_T$ of jets, leptons and 
$\mathbf{p}_T^{miss}$, and on the $\tau\tau$ invariant mass $m_{\tau\tau}$, calculated in the collinear approximation~\cite{massrec}
\begin{equation}
  | \mathbf{p}_T^{l_1} + \mathbf{p}_T^{l_2} + \mathbf{p}_T^{j} +\mathbf{p}_T^{miss} | < 30\,\textrm{GeV} \quad\quad 
  |m_{\tau\tau}-M_Z| > 25 \,\textrm{GeV}~.
\end{equation}
Events with identified b-jets (80\% efficiency, 6\% mistag) are rejected.
The WW control regions for the 0-jet and 1-jet bin are defined by omitting the $\Delta\phi_{\ell\ell}$ and $m_{\ell\ell,\,max}$
cuts of Eqs.~\eqref{phillcut} and \eqref{mllmaxcut} and requiring a minimal invariant lepton pair mass of
\begin{equation}
  m_{\ell\ell} > 80 \,\textrm{GeV} ~.
\label{mllcut}
\end{equation}
We do not consider the 2-jet channel, as there is not enough statistics at the moment 
for any conclusions in this channel.
Detector effects, efficiencies and hadronization effects have been neglected.

\section{Results for SUSY example}
\label{results}

\begin{figure}[tb]
  \begin{center}
    \includegraphics[height=0.33\textwidth]{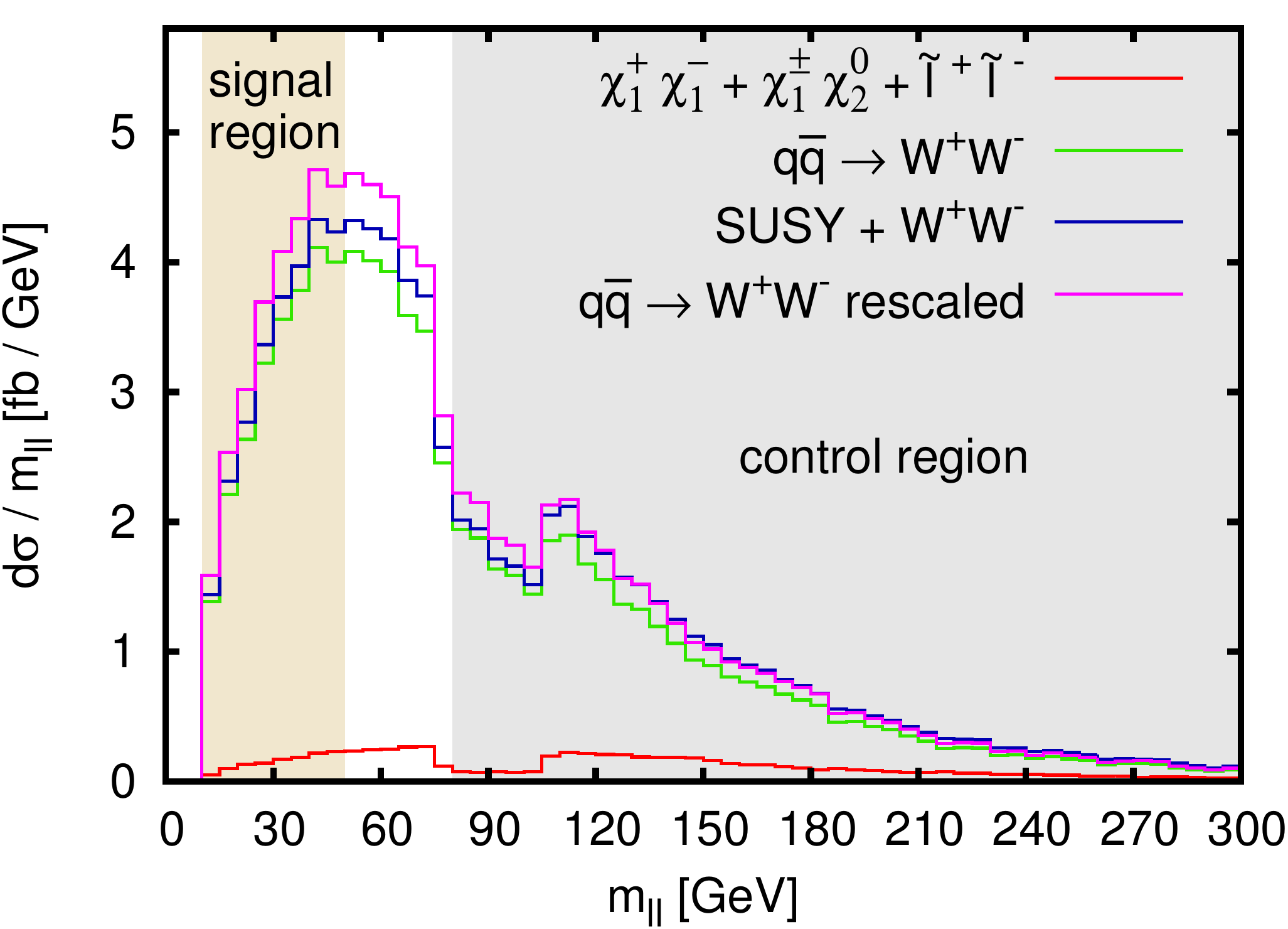}
    \hskip20pt
    \includegraphics[height=0.33\textwidth]{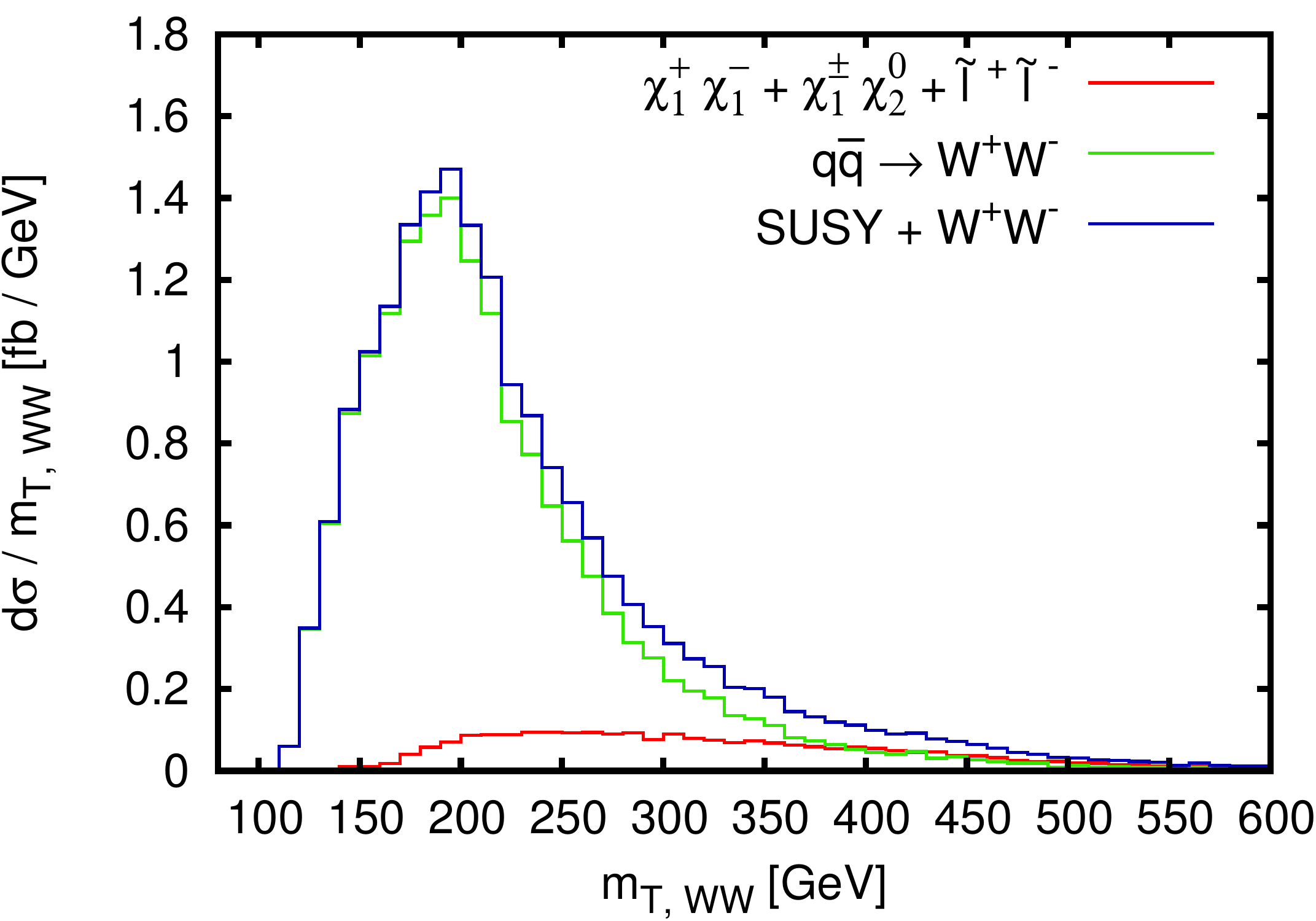}
  \end{center}
  \caption{Invariant lepton pair mass $m_{\ell\ell}$ (left) and transverse mass $m_T$ (right)
           distributions for the {\bf ``base'' scenario} of Section \ref{scenario}. The $m_T$
           distribution is calculated including all control region cuts of Eqs.~\eqref{basecuts} - \eqref{etmissrelcut},
           \eqref{ptllcut} and \eqref{mllcut}. For the $m_{\ell\ell}$
           plot the $m_{\ell\ell}$-cut of Eq.~\eqref{mllcut} is omitted.
           Both plots show the $q\bar{q}\rightarrow WW$ distribution, the SUSY contributions and their sum.
           The $m_{\ell\ell}$ plot also shows the $q\bar{q}\rightarrow WW$ result, rescaled by
           $(\sigma_C^{WW}+\sigma_C^{SUSY}) / \sigma_C^{WW}$, extracted from the control region.}
  \label{plots_basescenario}
\end{figure}

The effects of the SUSY processes of Eqs.~\eqref{ppx1x1} - \eqref{ppx1n2} on the full $m_{\ell\ell}$ range 
(signal\,\footnote{The full signal region cuts also include a cut on $\Delta\phi_{\ell\ell}$, Eq.~\eqref{phillcut}, 
but the effect on the $WW$ and SUSY processes is marginal in the $m_{\ell\ell} < 50\,\textrm{GeV}$ region.}
 and control region) in the 0-jet channel can be seen in Figure~\ref{plots_basescenario}.
The situation for the 1-jet channel is very similar,\footnote{Additional 
jets for the SUSY processes are simulated by the parton shower of {\tt Herwig++}.}
but the BSM effects are much less constrained compared to the 0-jet 
channel due to smaller event numbers. Therefore we focus on the 0-jet bin.
Chargino pair production accounts for the largest part of the SUSY signal, especially in the control region,
while slepton pair production has larger effects in the low $m_{\ell\ell}$
region due to the small assumed slepton masses.
As stated before, the production of a light chargino and the next-to-lightest neutralino has only a very small contribution
and is therefore not important for our results.
The relative contribution of the SUSY processes to the signal region is clearly much smaller than the
contribution to the control region. Therefore this scenario is potentially dangerous for the data-driven
estimation of the $q\bar{q}\rightarrow WW$ background: If the BSM physics could not be identified,
the $WW$ prediction in the signal region would be rescaled by 
\begin{equation}
 (\sigma_C^{WW}+\sigma_C^{SUSY}) / \sigma_C^{WW} \,,
\end{equation}
where $\sigma_C^{WW}$ and $\sigma_C^{SUSY}$ are the $WW$ and BSM contributions in the control region.
In our example, this leads to a $WW$ prediction for the signal region which is clearly too 
high (see Figure~\ref{plots_basescenario}).
Furthermore the effect on the shape of the $m_{\ell\ell}$ and $m_T$ distributions in the
signal region is too small for a detection of the SUSY contamination.

In contrast, a closer look at the transverse mass distribution can reveal
the BSM physics effects of this scenario. As $m_T$ is bounded from below by $m_{\ell\ell}$ and additional missing
transverse energy results in even larger values of $m_T$, BSM effects with large
$m_{\ell\ell}$ naturally lead to contributions at high $m_T$ values. This is especially the case in theories with
additional sources of missing energy like the MSSM.
Figure~\ref{plots_basescenario} shows an enhancement due to the SUSY contributions of more than 100\% for $m_T$ values
exceeding 350-400 GeV.
ATLAS measured 41 events with $m_T > 350\,\textrm{GeV}$, with a total background expectation of 
48 events, including a $WW$ contribution of 31 events \cite{atlasHinWW}. Therefore a factor of two increase in the
``$WW$ contribution'' is already ruled out with current data.

\begin{figure}[tb]
  \begin{center}
    \includegraphics[height=0.33\textwidth]{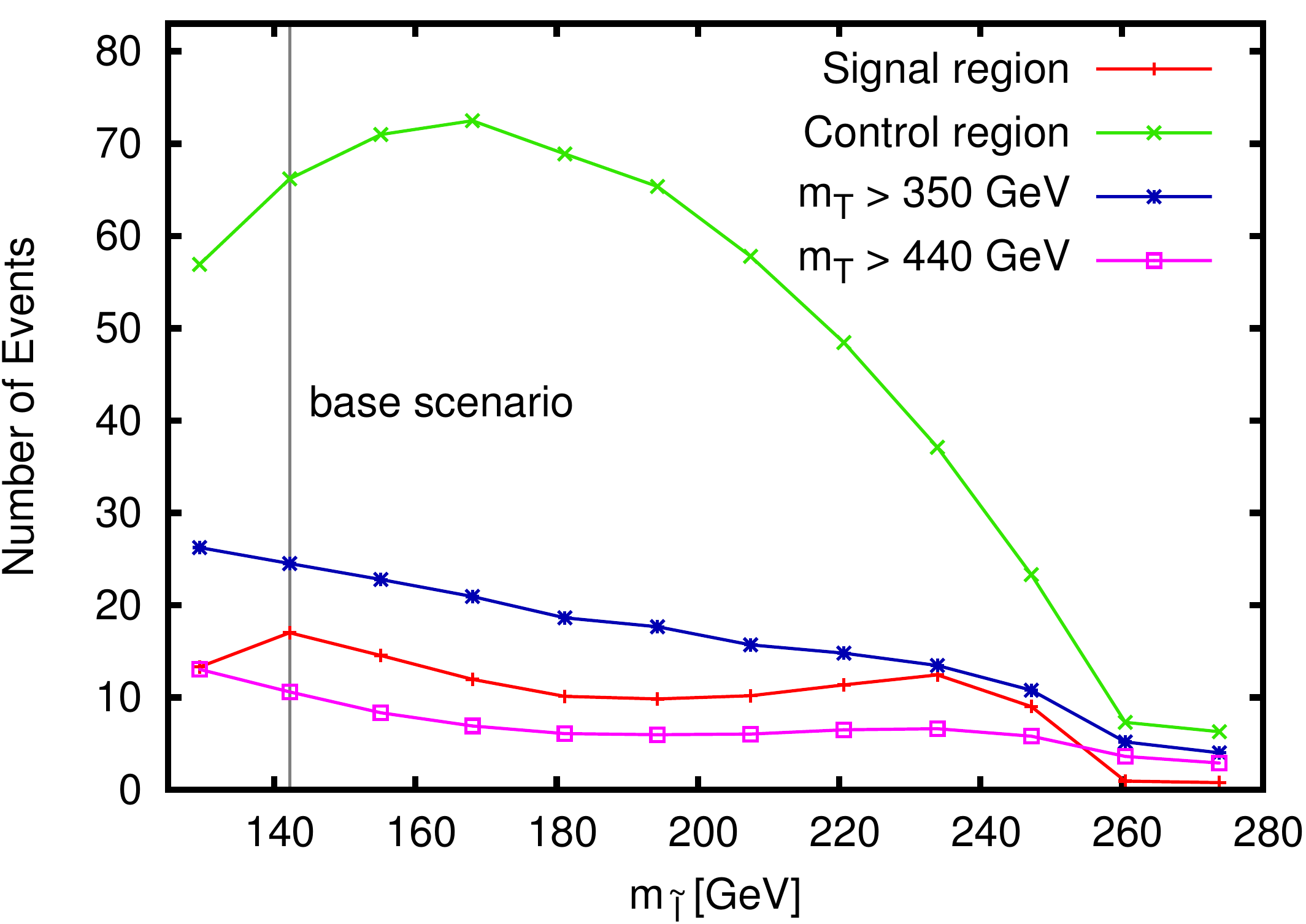}
    \hskip20pt
    \includegraphics[height=0.33\textwidth]{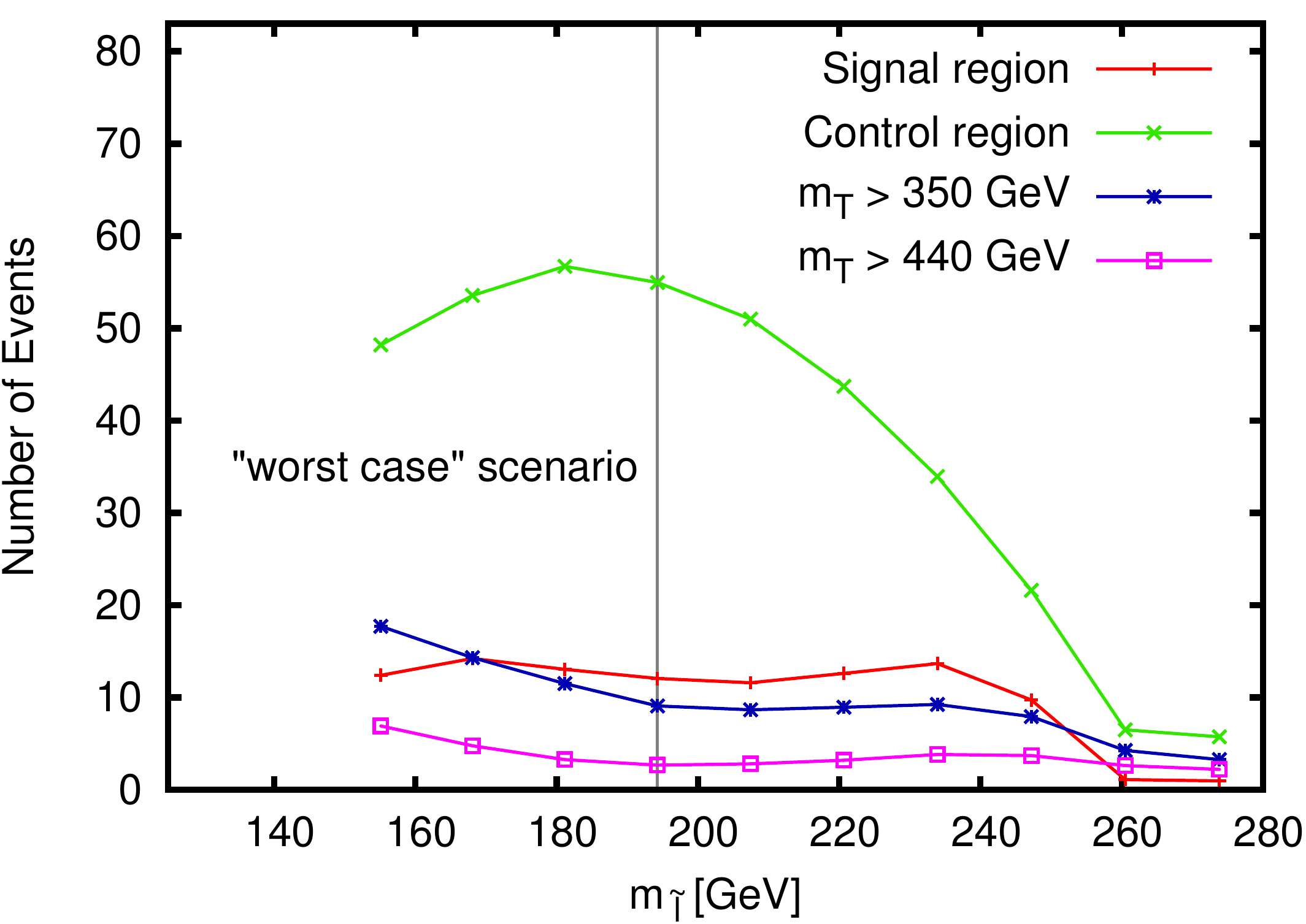}
  \end{center}
  \caption{Event numbers of the SUSY contributions in the signal region, in the control region and in the 
           control region with $m_T > 350\,\textrm{GeV}$ and $m_T > 440\,\textrm{GeV}$ for varying slepton
           masses of the first two generations. The LSP mass is $m_{\chi_1^0}=99\,\textrm{GeV}$
           in the left plot and $m_{\chi_1^0}=124\,\textrm{GeV}$ in the right plot. The discussed 
           ``base'' and ``worst case'' scenarios are marked.}
  \label{plots_series}
\end{figure}

Most of the SUSY contributions arise from chargino pair production. As the masses of the chargino decay products
play an important role for the kinematics of the final state leptons and for the amount of missing transverse momentum,
we vary the soft SUSY breaking parameters $M_1$, $M_{e L}$ and $M_{\mu L}$, which govern the LSP and left-handed
slepton masses.
The slepton mass variation also directly modifies the slepton pair production contributions to signal and
control region.
For the identification of a potentially dangerous scenario (concerning the normalization of the $WW$ background 
with the help of a control region) the following constraints have to be fulfilled:
\begin{itemize}
 \item The contribution to the signal region has to be as low as possible.
 \item The contribution to the control region has to be as large as possible,
       but still small enough to hide in the shape uncertainties of the control region.
 \item The part of the control region with high $m_T$ is strongly constrained by the current ATLAS data. Therefore
       the BSM effect in this region has to be small.
\end{itemize}

For the comparison with the ATLAS data from \cite{atlasHinWW}, 
we convert our cross sections into expected number of events by
normalizing our $WW$ prediction for $4.7 \,\textrm{fb}^{-1}$ with the expected
number of events from the ATLAS $H\rightarrow WW$ study within the 0-j control region.
From this rescaling we estimate an overall efficiency of 59\% for the evolution
of showered parton level events to reconstructed jets and leptons in the analysis.
For the rescaling we also take the $gg\rightarrow WW$ contribution into account using {\tt gg2WW}~\cite{gg2ww1,*gg2ww2},
which is included in the $pp \rightarrow WW$ background of the ATLAS study.
This part is formally of next-to-next-to leading order in QCD with respect to the
$q\bar{q}\rightarrow WW$ contribution and adds a few percent to the cross section~\cite{lhchiggsgroup}.
For the rest of our study, the $gg\rightarrow WW$ contribution has been neglected.

With this prescription the number of events shown in Figure~\ref{plots_series}
is calculated for a LSP mass of 99 GeV (left diagram)
and 124 GeV (right diagram). For each plot the slepton mass is varied up to the chargino mass, bounded from below 
by the requirement that the lightest neutralino has to be the LSP.
These event numbers have to be compared with the following values for the $q\bar{q}\rightarrow WW$ prediction
(our Monte Carlo prediction, scaled with the overall efficiency factor 0.59):
\begin{equation}
 N^{WW}_{signal} = 336 \quad\quad N^{WW}_{control} = 454 \quad\quad N^{WW}_{m_T>350\,\textrm{GeV}} = 22
 \quad\quad N^{WW}_{m_T>440\,\textrm{GeV}} = 7 \,,
\end{equation}
or $N^{WW}_{m_T>350\,\textrm{GeV}} = 31$ and $N^{WW}_{m_T>440\,\textrm{GeV}} = 11$ as taken from the ATLAS $m_T$ 
distribution~\cite{atlasHinWW}.
The discrepancy between those numbers can be ascribed to higher efficiencies for high $m_T$ events.
However, for our study we can largely eliminate this uncertainty by comparing ratios 
$N^{SUSY}_{MC} / N^{WW}_{MC}$ and the accuracy is sufficient for an approximate comparison with ATLAS data.

A slepton mass roughly in the middle between the LSP and chargino masses gives the largest contribution to
the control region. At the same time, the tail of the $m_T$ distribution is significantly smaller than for lighter
sleptons. Furthermore the larger slepton mass shifts the slepton pair production contribution from the signal
region to the control region.
The increased LSP mass in the right plot of Figure~\ref{plots_series} leads to less available kinetic energy
for the decay products and therefore to smaller $m_{\ell\ell}$, $m_T$ and $p_T^{miss}$.
This further reduces the contributions in the high $m_T$ bin of the control region.
We also considered a larger chargino mass as input. This reduces the chargino pair production cross section and 
therefore leads to a  reduction of the overall effect, as expected.

\begin{figure}[tb]
  \begin{center}
    \includegraphics[height=0.33\textwidth]{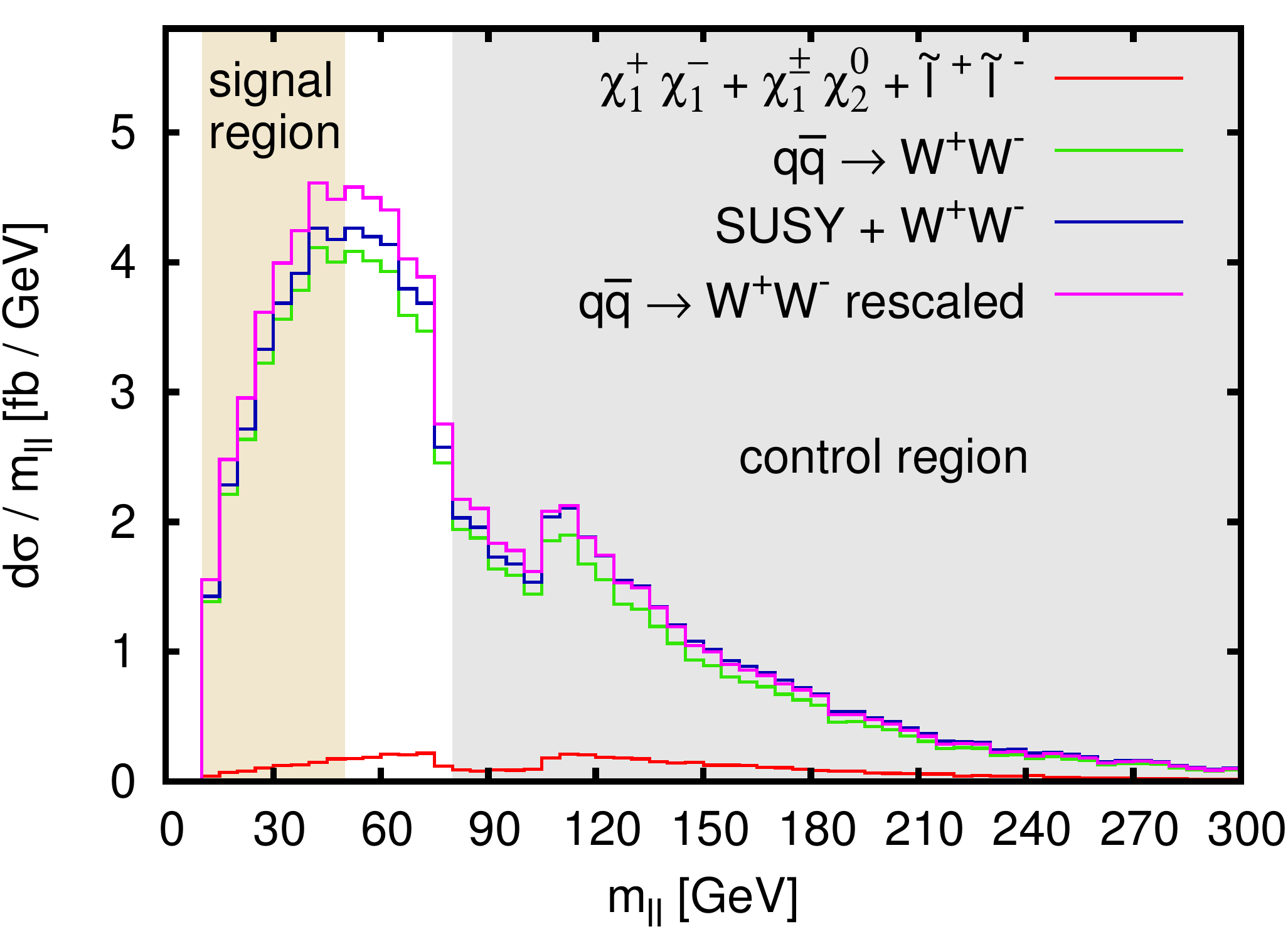}
    \hskip20pt
    \includegraphics[height=0.33\textwidth]{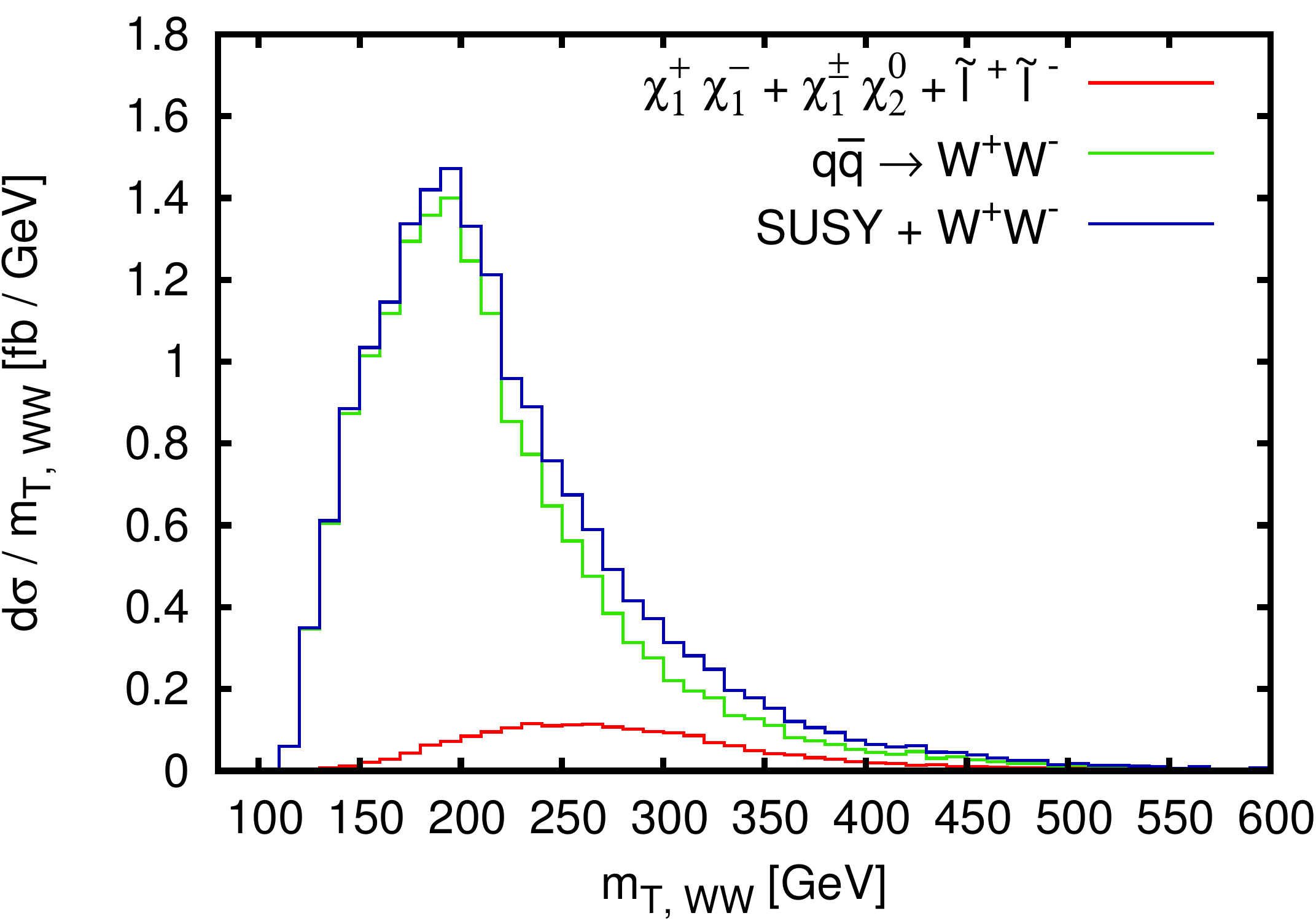}
  \end{center}
  \caption{Invariant lepton pair mass $m_{\ell\ell}$ (left) and transverse mass $m_T$ (right)
           distributions for the 
           {\bf ``worst case'' scenario} of Section \ref{results}.
           The $m_T$ distribution is calculated including all control region cuts of Eqs.~\eqref{basecuts} - \eqref{etmissrelcut},
           \eqref{ptllcut} and \eqref{mllcut}. For the $m_{\ell\ell}$
           plot the $m_{\ell\ell}$-cut of Eq.~\eqref{mllcut} is omitted.
           Both plots show the $q\bar{q}\rightarrow WW$ distribution, the SUSY contributions and their sum.
           The $m_{\ell\ell}$ plot also shows the $q\bar{q}\rightarrow WW$ result, rescaled by
           $(\sigma_C^{WW}+\sigma_C^{SUSY}) / \sigma_C^{WW}$, extracted from the control region.}
  \label{plots_worstcasescenario}
\end{figure}

Taking the criteria as described above, we find a scenario which is compatible with the
ATLAS $H\rightarrow WW$ analysis of 2011 data. We refer to it as the ``worst case'' scenario. 
As can be seen in Figures~\ref{plots_series} and~\ref{plots_worstcasescenario},
this scenario gives very small contributions to the high $m_T$ part of the
control region, small enough so that they cannot be identified
at the moment. At the same time the contributions to the signal region are
very small and therefore not noticeable,
although they are at partially higher values of $m_T$ than the $WW$ background.

We now want to quantify the effect of this BSM scenario on the $WW$ background prediction calculating the factor $C$
by which the expected number of $WW$ events in the signal region obtained from the normalization would have
to be corrected.
The extrapolation for the number of events from control to signal region is done using~\cite{lhchiggsgroup}
\begin{equation}
 N_{S} = \frac{N^{WW}_{S,MC}}{N^{WW}_{C,MC}} \, N_{C} = \alpha \cdot N_C \,.
\end{equation}
Taking both Standard Model $WW$ production and the BSM effects in the control region into account, this leads to
a predicted number of background events in the signal region given by
\begin{equation}
 N_{S}^{norm} = \alpha \cdot (N_C^{WW}+N_C^{SUSY}) \,,
\end{equation}
while the actual contribution is
\begin{equation}
 N_{S}^{true} = N_S^{WW}+N_S^{SUSY} \,.
\end{equation}
Therefore the predicted number of events would have to be reduced by
\begin{equation}
 C = \frac{N_S^{true}}{N_S^{norm}} = \frac{\sigma_S^{WW}+\sigma_S^{SUSY}}{\sigma_C^{WW}+\sigma_C^{SUSY}}
     \cdot \frac{\sigma_C^{WW}}{\sigma_S^{WW}}
\end{equation}
where we have replaced ratios of event numbers by our theoretical cross sections.
For this specific scenario we get
\begin{equation}
 C = 0.924 \,.
\end{equation}

For the ATLAS data which enters their fitting procedure and the resulting exclusion limits a selection 
cut of
\begin{equation}
 0.75 \cdot m_H < m_T < m_H
\end{equation}
on the transverse mass is applied.
Within this range the relative BSM contribution of our ``worst case'' scenario is even smaller,
leading to a larger correction for the extraction of the $WW$ background. In this case the number
of $WW$ events would have to be reduced by a factor of
\begin{equation}
 C = 0.897 \,.
\end{equation}
Since a SM Higgs signal is about 20\% of the overall background, an
overestimate of 10\% in the (dominant) $WW$ background would lead to a very
large underestimate in the size of the extracted Higgs signal.

\section{Conclusions}
\label{conclusion}
Data driven methods for background determination are extremely useful for
reducing theory errors inherent in QCD predictions for LHC cross
sections. However, they do rely on assumptions, namely that, apart from the
searched signal events, there are no other BSM contributions which could
affect the search. In this paper we have studied the impact of new physics contributions on the estimate of the SM 
background to the $H\to WW\to \ell\bar\nu \bar\ell\nu$ search.   In the case of W~pair production as
background to Higgs boson searches, the number of events is measured in a high
$m_{\ell\ell}$ control region where no signal events are to be expected. Via
an extrapolation using Monte Carlo predictions for the SM-shape of the
distributions of W~pair production, the estimate of the background in the 
softer signal region is determined.  In general, new physics at high energy
scales can enhance the number of events in hard control regions while
contributing little to a substantially softer signal region. The larger
measured rate in the control region then can lead to an overestimate of the
background in the signal region and dilute a potential signal.

 Two example scenarios in the context of the MSSM have been discussed in some detail. In the 
first one, the new physics contributions not only lead to an enhancement of
the event rates in the control region but also change the shape of
distributions (the $m_T$ distribution in the case considered) sufficiently to
make the extra BSM contributions noticeable.  The new 
physics contributions in the second example are more difficult to distinguish
from the pure SM case and could indeed have been missed in the
$H\rightarrow WW$ analyses. With enough data, one could, 
of course, see deviations in the hard event distributions, like in the tail 
of $m_T$ distributions for the scenarios at hand.

The MSSM scenario described above is just one example of BSM physics which
might affect the Higgs search and, as importantly, the measurement of Higgs
couplings from measured Higgs rates. Such potential BSM contamination should be
kept in mind when interpreting the Higgs search data within BSM scenarios: the
errors on Higgs boson couplings may be larger than in a pure SM analysis.

\section*{Acknowledgments}

We would like to thank Stefan Gieseke, Keith Hamilton and Christian R\"ohr
for discussions and {\tt Herwig++} support. We greatfully acknowledge helpful
discussions with Sophy Palmer, Eva Popenda and Michael Rauch. 
This work was supported by the BMBF under Grant No.~05H09VKG 
(\textquotedblleft Verbundprojekt HEP-Theorie\textquotedblright) 
and by the Initiative and Networking Fund of the Helmholtz 
Association, contract HA-101 (\textquotedblleft Physics at the Terascale\textquotedblright).

\begin{appendix}
\section{Parameters}
\label{parameter}
The ``base'' scenario is determined by the soft SUSY breaking parameters
\begin{equation}
\begin{array}{rclcrcl}
  M_1 &=& 103.1 \;\textrm{GeV} & \quad &
  M_{eL} = M_{\mu L} &=& 134.4 \;\textrm{GeV} \\
  M_2 &=& 270.1 \;\textrm{GeV} & \quad &
  M_{eR} = M_{\mu R} &=& 135.8 \;\text{GeV} \\
  M_3 &=& 1703.7 \;\textrm{GeV} & \quad &
  M_{\tau L} &=& 393.6 \;\textrm{GeV}  \\
  A_t &=& -2194.8 \;\textrm{GeV} & \quad &
  M_{\tau R} &=& 333.4 \;\text{GeV} \\
  A_b &=& -1907.2 \;\textrm{GeV} & \quad &
  M_{q_1 L} = M_{q_2 L} &=& 1579.8 \;\text{GeV} \\
  A_\tau &=& -249.4 \;\textrm{GeV} & \quad &
  M_{u R} = M_{c R} &=& 1524.3 \;\text{GeV} \\
  A_u = A_c &=& -655.5 \;\textrm{GeV} & \quad &
  M_{d R} = M_{s R} &=& 1517.7 \;\text{GeV} \\
  A_d = A_s &=& -821.8 \;\textrm{GeV} & \quad &
  M_{q_3 L} &=& 1201.4 \;\text{GeV} \\
  A_e = A_\mu &=& -251.1 \;\textrm{GeV} & \quad &
  M_{t R} &=& 1019.4 \;\text{GeV} \\
  M_{H_d}^2 &=& 32609 \;\textrm{GeV}^2 & \quad &
  M_{b R} &=& 1257.2 \;\text{GeV} \\
  M_{H_u}^2 &=& -169877 \;\textrm{GeV}^2 & \quad &
  \tan\beta(M_Z) &=& 10.0
\end{array}
\end{equation}
at the scale $Q=1\,\textrm{TeV}$
and the following Standard Model parameters, with the top mass from \cite{topmass}:
\begin{equation}
\begin{array}{rclcrcl}
  \alpha_{em}^{-1}(M_Z) &=& 127.934 & \quad &
  G_F &=& 1.16639 \cdot 10^{-5} \;\text{GeV}^{-2} \\
  \alpha_s(M_Z) &=& 0.1172 & \quad &
   M_Z &=& 91.187 \;\text{GeV} \\
   M_b(M_b) &=& 4.25 \;\text{GeV} & \quad &
   M_t &=& 173.2 \;\text{GeV} ~.
\end{array}
\end{equation}
These parameters are fed into {\tt SUSYHIT} \cite{SUSYHIT,*SuSpect,*SDECAY,*HDECAY} 
for the calculation of the SUSY particle masses and branching ratios. The SLHA output file is then used 
as input for {\tt FeynHiggs~2.8.6} \cite{FeynHiggs1,*FeynHiggs2, *FeynHiggs3, *FeynHiggs4}
in order to get precise Higgs boson mass values.
The resulting scenario exhibits the following features:
\begin{itemize}
\item The squark mass values of all three generations and the gluino mass values 
      ($m_{\widetilde{q}}\approx 1581\,\textrm{GeV}$, $m_{\widetilde{t}_1} = 934\,\textrm{GeV}$,
      $m_{\widetilde{b}_1} = 1232\,\textrm{GeV}$,
      $m_{\widetilde{g}} = 1725\,\textrm{GeV}$) are beyond current exclusion 
      limits~\cite{atlassusy1,*atlassusy2,cmssusy1,*cmssusy2,atlasstop,atlassbottom}. 
\item The trilinear coupling $A_t$ is adjusted according to the maximal mixing scenario~\cite{mhmax},
      which yields a Higgs boson mass with values in the vicinity of the experimental 
      hints of a Higgs boson~\cite{atlashiggscomb1,*atlashiggscomb2,cmshiggscomb1,*cmshiggscomb2}.
\item The wino mass parameter $M_2$ and the Higgs mass parameter $m_{H_u}$ are chosen to give chargino masses
      outside the exclusion limits of the ATLAS trilepton search \cite{atlascharginos}.
      The discussion of the $5\;\textrm{fb}^{-1}$ CMS trilepton analysis \cite{cmscharginos1,*cmscharginos2} 
      can be found in Section \ref{scenario}.
\item The stau lepton masses are larger than the light chargino mass 
      ($m_{\widetilde{\tau}_1} = 334\,\textrm{GeV}$), which is a specific feature of the considered scenarios.
\item The mass parameters for the left-handed sleptons of the first two generations $M_{e L}$ and 
      $M_{\mu L}$ are chosen such that the chargino decay into selectrons and smuons is the dominant decay mode.
\end{itemize}
The soft SUSY breaking parameters of the ``worst case'' scenario discussed in Section \ref{results}
are the same as in the ``base'' scenario, except for
\begin{equation}
\begin{array}{rclcrcl}
  M_1 &=& 128.9 \;\textrm{GeV} & \quad &
  M_{eR} = M_{\mu R} &=& 190.1 \;\text{GeV} \\
  M_{eL} = M_{\mu L} &=& 188.2 \;\textrm{GeV} \;.
\end{array}
\end{equation}
The relevant masses and branching ratios of both scenarios are listed in Table \ref{masstab}.

\tabcolsep1.9mm
\begin{table}[tb]
  \begin{center}
  \begin{tabular}{ccc}
  \toprule
   & ``base'' scenario & ``worst case'' scenario \\
  \midrule
   & $M_{e,\mu\, L,R}=M_{e,\mu\, L,R}^0$, & $M_{e,\mu\, L,R}=1.4 \cdot M_{e,\mu\, L,R}^0$,\\
   & $M_1=103\,\textrm{GeV}$ & $M_1=129\,\textrm{GeV}$ \\
  \midrule
  \midrule
  $m_{\chi_1^0}$                            &  98.9 GeV &  124.1 GeV  \\
  $m_{\chi_1^+}$                            & 260.0 GeV & 260.3 GeV  \\
  $m_{\chi_2^0}$                            & 260.3 GeV & 260.7 GeV  \\
  \midrule
  $m_{\widetilde{e}_L}=m_{\widetilde{\mu}_L}$ & 141.8 GeV & 193.5 GeV \\
  $m_{\widetilde{e}_R}=m_{\widetilde{\mu}_R}$ & 142.6 GeV & 195.0 GeV \\
  \midrule
  $BR(\chi_1^+\rightarrow \ell^+\,\widetilde{\nu}_\ell)$   & 58.0 \% & 60.4 \% \\
  $BR(\chi_1^+\rightarrow \widetilde{\ell}^+_L\,\nu_\ell)$ & 41.0 \% & 37.7 \% \\
  $BR(\chi_1^+\rightarrow W^+\,\chi_1^0)$                  &  1.0 \% &  1.9 \% \\
  \midrule
  $BR(\chi_2^0\rightarrow \widetilde{\ell}_{L,R}^\pm\,\ell^\mp)$ & 47.7 \% & 45.1 \% \\
  $BR(\chi_2^0\rightarrow \widetilde{\nu}_\ell\,\bar{\nu}_\ell)$ & 51.4 \% & 53.2 \% \\
  $BR(\chi_2^0\rightarrow \chi_1^0 \,Z)$                         &  0.2 \% &  0.3 \% \\
  $BR(\chi_2^0\rightarrow \chi_1^0 \,h_0)$                       &  0.7 \% &  1.4 \% \\
  \midrule
  $BR(\widetilde{\ell}_{L,R}^\pm\rightarrow \chi_1^0 \, \ell^\pm)$ & 100.0\% & 100.0\% \\
  \bottomrule
  \end{tabular}
  \end{center}
  \caption{Masses and branching ratios of interest for our ``base'' scenario and
           for a scenario with modified $M_1$ and $M_{e,\mu\, L,R}$.
           Here, $\stackrel{\text{\tiny(}\sim\text{\tiny)}}{\ell}$ means 
           the combined (s)electron and (s)muon channel. For completeness, we also give 
           the mass and the branching ratios of the $\chi_2^0$ though they are not important for our results.}  
  \label{masstab}
\end{table}

\end{appendix}

\bibliography{SUSYinControlWW}

\end{document}